\documentclass[prl,aps, twocolumn, superscriptaddress,twocolumn,notitlepage]{revtex4-2}

\def\be{\begin{equation}}
\def\ee{\end{equation}}
\usepackage{graphicx}
\usepackage{dcolumn}
\usepackage{bm}
\usepackage{float}
\usepackage{amsmath}
\usepackage{amssymb}
\usepackage{lineno}
\usepackage[bookmarks=true,
   colorlinks=true,
   linkcolor=blue,
   urlcolor=blue,
   citecolor=blue,
   bookmarks=true,
   hyperindex=true
]{hyperref}
\usepackage{natbib}
\usepackage{orcidlink}

\usepackage{soul}
\newcommand{\SHS}[1]{\textcolor{blue}{#1}}

\begin{document}

\title{Reversibility of Nonequilibrium Dynamics in a Unitary Fermi Gas}

\author{Xi-Zhi Wu}
\affiliation{Innovation Academy for Precision Measurement Science and Technology, Chinese Academy of Sciences, Wuhan 430071, China}
\affiliation{University of the Chinese Academy of Sciences, Beijing 100049, China}

\author{Xiangchuan Yan}
\affiliation{Innovation Academy for Precision Measurement Science and Technology, Chinese Academy of Sciences, Wuhan 430071, China}

\author{Xin Xie}
\affiliation{Innovation Academy for Precision Measurement Science and Technology, Chinese Academy of Sciences, Wuhan 430071, China}
\affiliation{University of the Chinese Academy of Sciences, Beijing 100049, China}

\author{Dali Sun}
\email{dlsun@wipm.ac.cn}
\affiliation{Innovation Academy for Precision Measurement Science and Technology, Chinese Academy of Sciences, Wuhan 430071, China}

\author{Seong-Ho Shinn\orcidlink{0000-0002-2041-5292}}
\affiliation{Department of Physics and Materials Science, University of Luxembourg, L-1511 Luxembourg, Luxembourg}

\author{Adolfo del Campo\orcidlink{0000-0003-2219-2851}}\email{adolfo.delcampo@uni.lu}
\affiliation{Department of Physics and Materials Science, University of Luxembourg, L-1511 Luxembourg, Luxembourg}
\affiliation{Donostia International Physics Center, E-20018 San Sebastian, Spain}

\author{Kaijun Jiang\orcidlink{0000-0003-3126-4335}}\email{kjjiang@wipm.ac.cn}
\affiliation{Innovation Academy for Precision Measurement Science and Technology, Chinese Academy of Sciences, Wuhan 430071, China}

\date{\today}

\begin{abstract}
We theoretically describe and experimentally observe the reversibility of a nonequilibrium state in a unitary Fermi gas confined in a time-dependent trap. Unlike the equilibrium case, reversibility out of equilibrium imposes a stringent condition: the accumulated dynamical phase must be an integer multiple of $2\pi$, i.e., $\Delta\phi = 2\omega_{0} s(T) = 2n\pi$, where $\omega_{0}$ is the initial trapping frequency and $s(T)$ is the rescaled time after a total evolution time $T$. We demonstrate the reversibility of an initial nonequilibrium state in an isotropic unitary Fermi gas by driving the optical trap through a round trip.  We find two types of reversible pathways, the space-time scaling-invariant and polynomial shortcut-to-adiabaticity (STA) protocols. The phase condition restricts the round-trip time $T$ to discrete values, a distinguishing feature of nonequilibrium dynamics. Our work establishes a clear criterion for the quantum control of nonequilibrium many-body systems beyond linear response and provides practical pathways to reversibility.
\end{abstract}

\maketitle

\emph{Introduction---} Scale invariance is a powerful symmetry that constrains the real-time dynamics of quantum many-body systems. In ultracold atom physics, it has applications ranging from imaging \cite{Castin1996,Kagan1997} and microscopy \cite{delcampo2011,Papoular2015} to quantum control \cite{Muga2009,Schaff2011,delCampo2012,delcampo2013,delCampo2021} and quantum thermodynamics \cite{Jaramillo2016,Beau2016,Deng2018SciAdv}. When present, it equips the Hamiltonian with an SO(2,1) algebra, whose generators relate states of different energies and yield exact results for collective observables such as the mean-square radius \cite{Pitaevskii1997,Werner2006PRA,Castin2004}, mean energy \cite{Jaramillo2016,Beau2016,Deng2018SciAdv},  energy fluctuations \cite{Beau2020,delCampo2021,Peng2026}, and the survival probability \cite{delcampo2016,Fogarty2020,delCampo2021,Yang2025}. 
The three-dimensional (3D) unitary Fermi gas, a spin-$1/2$ Fermi mixture tuned to a broad Feshbach resonance, is a preeminent realization of this symmetry. The divergent $s$-wave scattering length $a\to\infty$ eliminates the only microscopic length scale, so that the contact interaction exhibits scale invariance $V(\Lambda\bm{r})=\Lambda^{-2}V(\bm{r})$ \cite{Chin2010}. Confined in an isotropic harmonic trap, the gas therefore possesses an exact dynamical SO(2,1) symmetry \cite{Pitaevskii1997,Werner2006PRA,ZhoufeiPRA2019Conformal,ZhoufeiPRA2020Conformal,Zhou2020PRLgeometrizing,Zhou2020PRLecho}.

Scale invariance has striking dynamical consequences. In free expansion, the normalized cloud size grows as an exact power law, $\langle x_i^2\rangle(t)/\langle x_i^2\rangle_0 = 1+\omega^{2}_0 t^2$ where $\omega_0$ is the trapping frequency before the expansion \cite{Castin2004,OHara2002,Elliott2014}. This scale-invariant expansion has been quantitatively confirmed in an isotropic trap \cite{Wang2024}, and the related Efimovian and super-Efimovian expansion dynamics have been observed in scale-invariant Fermi gases \cite{Deng2016Science,Deng2018PRL}. Within the trap, the SO(2,1) algebra forces the breathing mode to oscillate at twice the trapping frequency $\omega_B=2 \omega_0$ and with nearly zero damping rate $\Gamma_B\to 0$ \cite{Pitaevskii1997,Werner2006PRA,Wang2024,Sun2025}. 
While the equilibrium and linear-response properties of scale-invariant gases are well understood, comparatively little is known about the reversibility of nonequilibrium driving protocols when the initial state is itself far from equilibrium.  For thermalized states, shortcuts to adiabaticity (STA) provide a powerful family of protocols that transfer the system between two equilibrium states in finite time without residual excitations \cite{GuerryOdelin2019,Chen2010,Deffner2014}.  These methods have been applied to thermal clouds, 
Bose-Einstein condensates (BECs),
and Fermi gases in the noninteracting and strongly-interacting regimes, including the suppression of quantum friction via superadiabatic protocols \cite{Deffner2014,Deng2018,Diao2018,Deng2018SciAdv}. However, the extension to initially nonequilibrium states requires qualitatively new considerations because the internal quantum state carries the phase information of its internal oscillation, which conventional STA boundary conditions do not constrain. Driving the control parameter through a cycle provides a natural way to assess reversibility and adiabaticity in this context.

In this work, we explore the conditions for the reversibility of nonequilibrium states in a 3D isotropic unitary Fermi gas. Starting from an initial nonequilibrium state associated with a breathing mode, we identify three independent conditions that the round-trip scan of the trap frequency must satisfy for the dynamics to be reversible: (i) the scaling factor must return to unity, $b(T)=1$; (ii) its time-derivative must vanish, $\dot{b}(T)=0$; and (iii) the accumulated dynamic phase must be an integer multiple of $2\pi$, $\Delta \phi = 2\omega_{0}s(T)=2n\pi$, where $s(T)$ is the rescaled internal time over the round-trip time $T$. Conditions (i) and (ii) are satisfied by conventional STA designed for thermalized states, but condition (iii) is generically violated because different protocol paths accumulate different dynamic phases. We experimentally provide two kinds of pathways, the 
space-time 
scale-invariant (inversely proportional trap manipulation $\omega(t)\propto t^{-1}$) and polynomial STA protocols, to observe the reversibility. The phase condition (iii) restricts the round-trip time $T$ to be discretized, exhibiting the unique feature for the nonequilibrium dynamics.

\emph{Conditions for reversibility---} We consider $N$ fermionic atoms of mass $m$ and two spin states $\{\uparrow,\downarrow\}$ confined in a 3D isotropic harmonic trap of instantaneous frequency $\omega(t)$.
The Hamiltonian reads
\begin{equation}
\hat{H}(t) = \sum_{i=1}^{N}\frac{-\hbar^2\nabla_i^2}{2m}+\sum_{i = 1, j > i}^N\! V(\bm{r}_i-\bm{r}_j)+ \frac{m\omega^2(t)}{2}\sum_{i=1}^{N}r_i^2,
  \label{eq:H}
\end{equation}
where the pairwise interaction $V$ is the zero-range potential characterized by the scattering length $a$. At unitarity, $a\rightarrow\infty$, the potential satisfies $V(\Lambda\bm{r})=\Lambda^{-2}V(\bm{r})$ \cite{Werner2006PRA}. The full Hamiltonian at fixed $\omega$ obeys the scaling relation \cite{Deng2016Science}, $ \hat{H}\!\left(\bm{r} \to \Lambda\bm{r},\,\omega\to\Lambda^{-2}\omega\right) = \Lambda^{-2}\, \hat{H} (\bm{r},\omega)$, which is the basis of the SO(2,1) dynamical symmetry.

When the trap frequency $\omega(t)$ is time-dependent, the $N$-body wavefunction evolves in a \emph{self-similar} fashion \cite{Werner2006PRA,Gritsev2010,delcampo2011,delCampo2021,Dupays2021,Dupays2023}.  Specifically, if $\Psi_0\equiv\Psi(t=0)$ is any eigenstate of $H(0)$, then the exact time-evolved state is
\begin{equation} \label{eq:psi}
\begin{split}
  \Psi(\{\bm{r}_i\};t) &= \frac{1}{ b^{3 N / 2} \left( t \right)}\exp\!\left[\frac{i m \dot{b}(t)}{2\hbar b(t)}\sum_{i=1}^N r_i^2\right] \\
  &\times \Phi\!\left(\!\left\{\frac{\bm{r}_i}{b(t)}\right\};\,s(t)\right),
\end{split}
\end{equation}
where $b(t)>0$ is the time-dependent scaling factor describing the collective dilation of the cloud. The prefactor $b^{-3N/2}\left( t \right)$ ensures normalization. The exponential is the chirp phase, a quadratic function of position that encodes the residual radial velocity field of the expanding/contracting cloud. The wavefunction $\Phi(\{\bm{u}_i\};s)$ is the internal state in the co-moving frame, evolving in the rescaled time  
 \begin{equation}
 s(t) \equiv \int_0^t \frac{dt'}{b^2 \left( t' \right)} .
 \label{eq:rescaled_time}
 \end{equation}
The internal state satisfies the unperturbed Schr\"odinger equation, $i\hbar\,\partial_s\,\Phi(\{\bm{u}_i\};s) = \hat{H}_0 \,\Phi(\{\bm{u}_i\};s) $, where $\hat{H}_0 \equiv \hat{H} \left( t = 0 \right)$ is the initial Hamiltonian with frequency $\omega_{0}$. 
This constrains $b(0) = 1$ and $\dot{b}(0) = 0$. 
The internal quantum state is decoupled from the global dilation: it simply precesses under $\hat{H}_0$ at the rate set by the accumulated rescaled time $s(t)$.


The cloud observable directly measured in experiments is the mean-square radius $\langle{r^2}\rangle(t)=\sum_{i = 1}^N \langle{r_i^2}\rangle(t)$. Using Eq. \eqref{eq:psi} in the time-dependent Hamiltonian of Eq. \eqref{eq:H}, one obtains the exact factorization (see Supplemental Material \cite{Supplemental_Reversibility} for details)
\begin{equation} \label{eq:factorisation}
  \langle{r^2}\rangle(t) = b^2 \left( t \right) \langle{r^2}\rangle_\Phi(s(t)),
\end{equation}
where
\begin{eqnarray}
  \ddot{b} \left( t \right) + \omega^2 \left( t \right)b \left( t \right) = \frac{\omega_{0}^2}{b^3 \left( t \right) }, \label{eq:ermakov}\\
  \langle{r^2}\rangle_\Phi(s) = 
\langle{r^2}\rangle_{0}[1 + A_m\cos(2\omega_{0} s + \phi_{0})], \label{eq:internal_oscillation}
\end{eqnarray}
with the oscillation amplitude $A_m$, phase $\phi_{0}$, and balanced size $\langle{r^2}\rangle_{0}$
inherited from the initial state. 
Equation \eqref{eq:factorisation} is key in our analysis: the total cloud size separates into a geometric envelope $b^2 \left( t \right)$ which is entirely controlled by the protocol through the Ermakov equation \eqref{eq:ermakov}, and an internal oscillation $\langle{r^2}\rangle_\Phi(s(t))$ whose phase accumulates at the rate $2\omega_{0} \dot{s}=2\omega_{0}/ b^2 \left( t \right)$. 

As a quantum echo, we now consider a round-trip protocol of total duration $T$: the trap frequency $\omega(t)$ is modulated from $\omega_{0}$ to some intermediate point and back \cite{Supplemental_Reversibility}, and we ask under what conditions the state $\Psi(\{r_i\};T)$ is identical to $\Psi(\{r_i\};0)$.  

The geometric factor in Eq. \eqref{eq:factorisation} must return to its initial value and the chirp phase in Eq. \eqref{eq:psi} introduces a position-dependent phase difference between $\Psi(T)$ and $\Psi(0)$ \cite{Dupays2021}. Full state restoration requires
\begin{equation}
b(T) = 1,  \tag{C-I} \label{eq:C1} 
\end{equation}
\begin{equation}
\dot{b}(T) = 0 . \tag{C-II} \label{eq:C2}
\end{equation}
Together, conditions \eqref{eq:C1} and \eqref{eq:C2} define the family of ``STA protocols for thermalized states'' \cite{Chen2010,delcampo2011,delcampo2013,Deng2018,GuerryOdelin2019}, and they are sufficient for the reversibility of any initial equilibrium state.

However, for a nonequilibrium initial state, conditions \eqref{eq:C1} and \eqref{eq:C2} are necessary but not sufficient. The internal oscillation in Eq. \eqref{eq:internal_oscillation} may accumulate a phase shift relative to the undriven case. Full restoration of the state further requires $\Phi(s(T))=\Phi(s(0))$, which demands
\begin{equation}
\Delta\phi = 2\omega_{0}\,s(T) = 2n\pi, \qquad n\in\mathbb{Z}. \tag{C-III} \label{eq:C3}
\end{equation}
Since $s(T)$ depends on the entire trajectory of $b(t)$, not just its endpoints, different protocols that satisfy \eqref{eq:C1} and \eqref{eq:C2} will generically fail \eqref{eq:C3}. The condition \eqref{eq:C3} is absent for thermalized initial states and universal for any nonequilibrium state within the SO(2,1) symmetry (all such states oscillate at $2\omega_{0}$ in rescaled time). The breathing-mode phase thus acts as a protocol-discriminating observable.

Next, we focus on the experimental observation of reversibility of an initial nonequilibrium state. Our experiments are performed with a cloud of $^6$Li atomic degenerate Fermi gas with two equal spin states $\vert F=1/2,m_F = \pm 1/2 \rangle$ in a 3D isotropic dipole trap \citep{Wang2024, Sun2025, Yan2026}. The atomic cloud typically contains $N=7.5(2)\times10^4$ atoms per spin state at a temperature $T/T_F=0.27(1)$, where $T_F$ is the Fermi temperature of an ideal Fermi gas. The magnetic field is set to 832.2 G, right on the broad Feshbach resonance. The asphericity of the trap is smaller than $5\%$. 
Two kinds of pathways, the 
space-time scale-invariant and polynomial STA protocols, are applied in the experiment.

\begin{figure}
\includegraphics[width=8.5 cm]{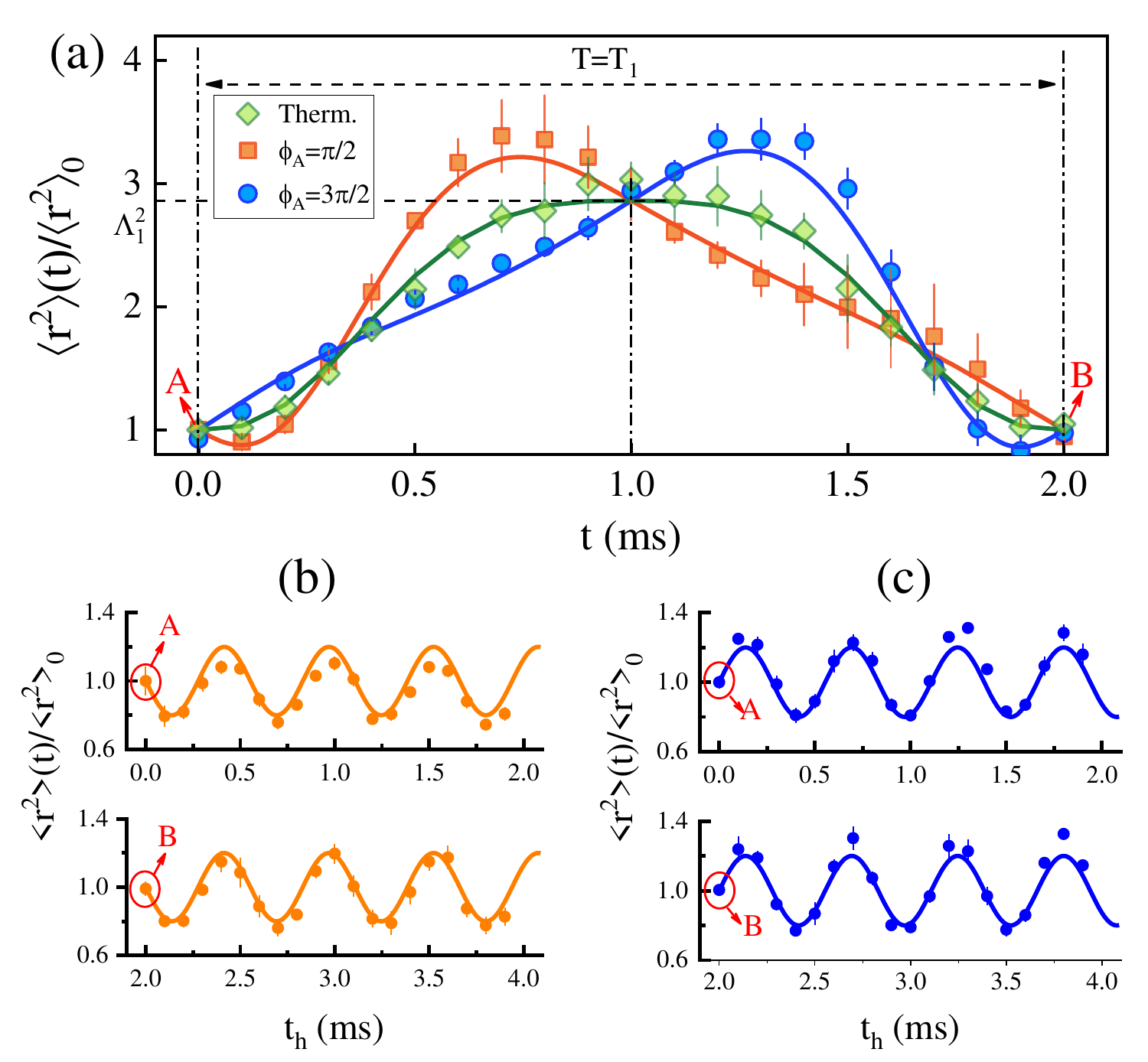}
\caption{Space-time scale-invariant protocol for 
reversibility. Here, $n = 1$ in Eq. \eqref{eq:scaling period}. With $\lambda = 0.109$, the round-trip time is $T=T_1 \approx 2 \ \textrm{ms}$. (a) Normalized atomic cloud sizes $\langle{r^2}\rangle(t)/\langle{r^2}\rangle_{0}$ during the trap scan. $A$ and $B$ denote the start and end points of the trap manipulation, respectively. An initially thermalized atomic cloud (green diamonds) is shown as a reference, and at the midpoint, it reaches the first plateau with $\Lambda_{1}^{2} = 2.86$. The phases of the two initial nonequilibrium systems are set as $\phi_{A}=\pi/2$ (orange squares) and $\phi_{A}=3\pi/2$ (blue circles). (b) Breathing-mode oscillations at $A$ and $B$ points for the nonequilibrium system with $\phi_{A}=\phi_{B}=\pi/2$. Here the trapping frequency is kept constant as $\omega(t)=\omega_0$ during the trap-holding time $t_h$. (c) Nonequilibrium system with $\phi_{A}=\phi_{B}=3\pi/2$. The solid curves are the calculations of Eq. \eqref{eq:factorisation}. Error bars represent the standard deviation of several measurements.} \label{Fig1}
\end{figure}

\begin{figure}
\includegraphics[width=8.5 cm]{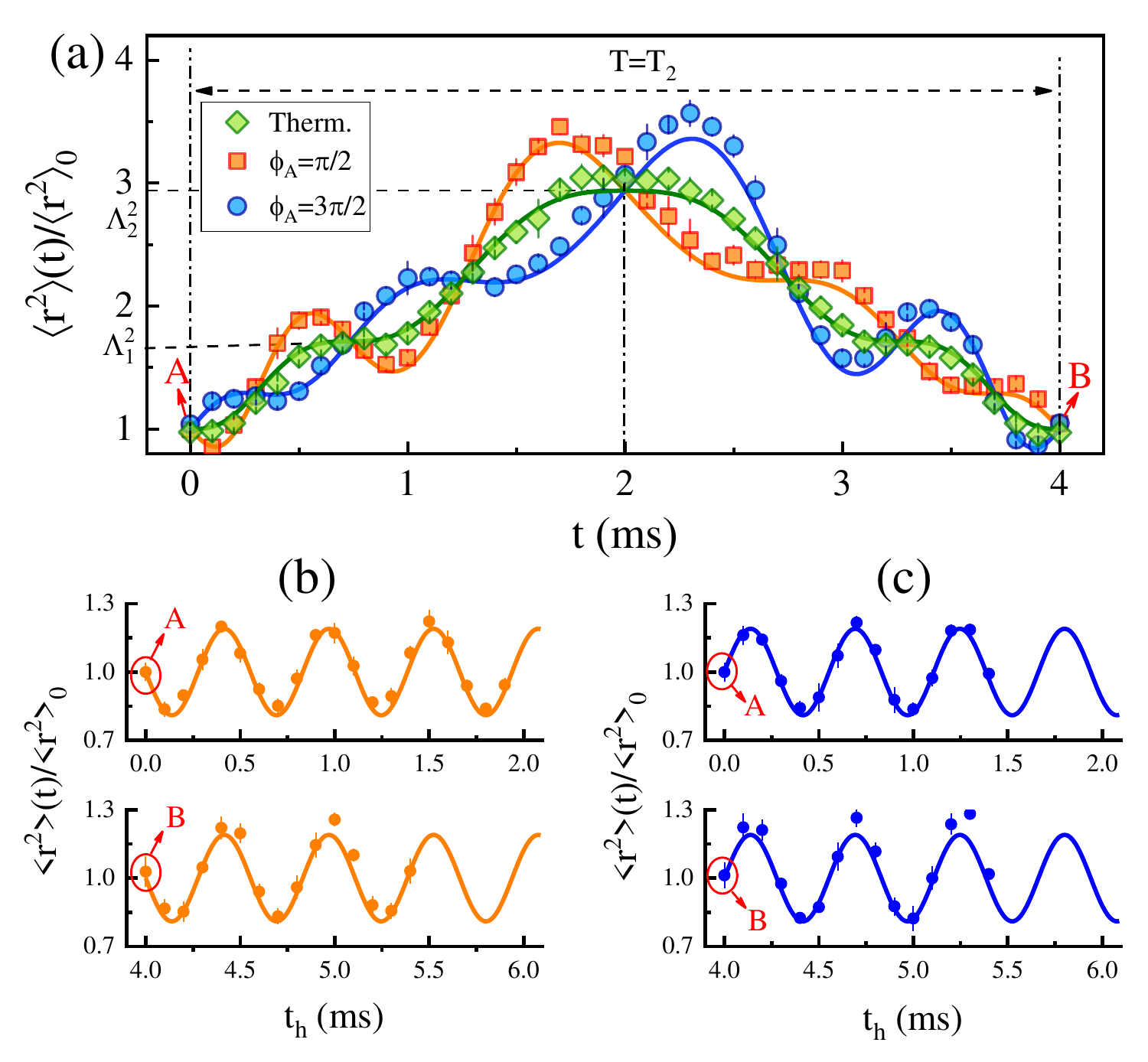}
\caption{ Space-time scale-invariant protocol for 
reversibility, paralleling the presentation in Fig. \ref{Fig1}. Here, $n = 2$ in Eq. \eqref{eq:scaling period}. With $\lambda = 0.029$, the round-trip time is $T=T_2 \approx 4 \ \textrm{ms}$. At the midpoint $t=T/2=2\ \textrm{ms}$, the initially thermalized atomic cloud reaches the second plateau with $\Lambda_{2}^{2} = 2.93$. The phases of the two initial nonequilibrium systems recover at the end point of the trap manipulation, $\phi_{B} \approx \phi_{A}=\pi/2\ (\textrm{or}\ 3\pi/2)$.} \label{Fig2}
\end{figure} 

\emph{Space-time scale-invariant protocol---} We scan the trapping frequency inversely proportional to time as follows:
\begin{equation} \label{eq:scaling scan}
\omega(t) = \begin{cases}
\dfrac{1}{\sqrt{\lambda}(t_0+t)}, &  0 \leq t \leq T/2, \\
\dfrac{1}{\sqrt{\lambda}(T+t_0-t)}, & T/2 \leq t \leq T, 
\end{cases}
\end{equation}
where $\omega_0=1/(\sqrt{\lambda}t_0)$ is the initial trapping frequency, $T$ is the round-trip time, and $\lambda$ characterizes the speed of the trap change. Under this variation of the trap, the system exhibits 
space-time scale invariance, i.e., $\hat{H}\!\left(\bm{r} \to \Lambda \bm{r}, t \to \Lambda^{2}t\right) = \Lambda^{-2}\, \hat{H} (\bm{r},t)$. For $\lambda < 4$, the initial thermalized atoms develop discrete plateaus \cite{Deng2016Science}, and the trajectory from the initial state to these plateaus forms accidental STA paths, i.e., $b(T)=1$ and $\dot{b}(T)=0$. 
In this experiment, we set the system to reach the $n$-th plateau at the midpoint $t=T_n/2$ with $T = T_n$. This leads to 
\begin{equation} \label{eq:scaling period}
T_{n}/2=t_0(e^{2n\pi/\alpha_0}-1), \ n\in \mathbb{Z},
\end{equation}
where $\alpha_0=2\sqrt{1/\lambda-1/4}$.
Based on the trap trajectory of Eq. \eqref{eq:scaling scan}, we calculate the accumulated dynamic phase during the round trip,
\begin{equation} \label{scaling-phase}
\Delta \phi =2\omega_0 s(T)=4n\pi, \  n\in \mathbb{Z},
\end{equation}
which automatically satisfies the phase condition \eqref{eq:C3} (see calculation details in Supplemental Material \cite{Supplemental_Reversibility}).

\begin{figure*}
\includegraphics[width=17 cm]{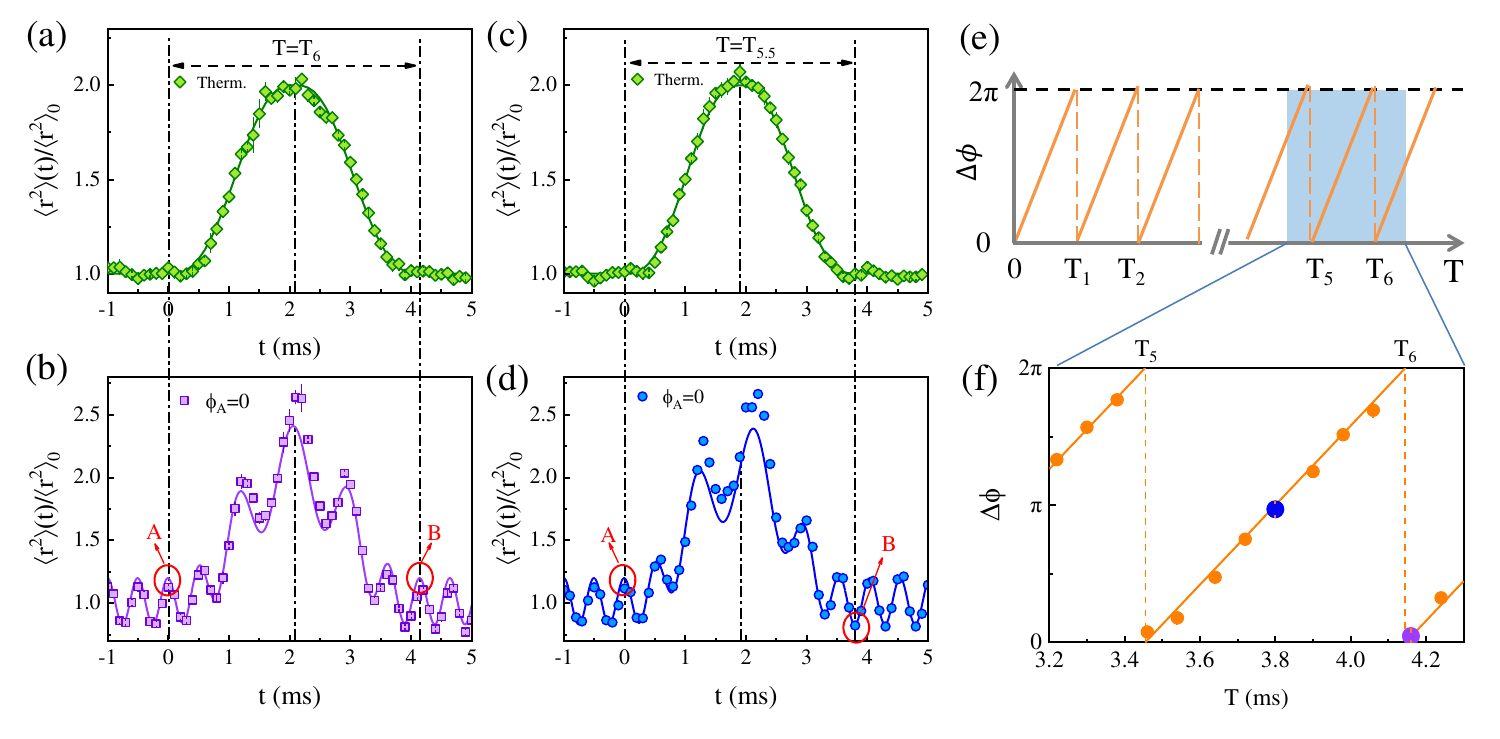}
\caption{Polynomial STA protocol for reversibility. Panels (a) and (b) show the normalized atomic cloud sizes $\langle{r^2}\rangle(t)/\langle{r^2}\rangle_{0}$ during the trap manipulation. Here $n = 6$ in Eq. \eqref{polynomial time} and the round-trip time is $T_6 = 4.16\ \textrm{ms}$. For $t<0$ and $t>4.16\ \textrm{ms}$, the trapping frequency is kept at $\omega(t)=\omega_0$ to measure the breathing-mode oscillation. For $0 \leq t \leq 4.16\ \textrm{ms}$, the trapping frequency is scanned with a round trip, where $A$ and $B$ denote the start and end points of the scan, respectively. In (a), an initially thermalized atomic cloud is shown as a reference. In (b), the phase of the initial nonequilibrium system is set as $\phi_{A}=0$. At the end point, the phase is $\phi_{B}=0$. Panels (c) and (d) show the measurements for $n = 5.5$, where  the round-trip time is $T_{5.5} = 3.80\ \textrm{ms}$. For the initially thermalized atomic cloud in (c), it is reversible. For the initial nonequilibrium system with $\phi_{A}=0$ in (d), the phase at the end point is $\phi_{B}=\pi$, obviously different from that at the start point. 
Panel(e) schematically shows the phase difference, $\Delta\phi=\phi_{B}-\phi_{A}$, as a function of the round-trip time $T$. 
Panel (f) shows the measurements of $\Delta\phi$ for different $T$, where the purple and blue circles represent the measurements in (b) and (d), respectively. The orange solid curve denotes the calculation of Eq. \eqref{polynomial phase}. Error bars represent the standard deviation of several measurements.} \label{Fig3}
\end{figure*}

In Fig. \ref{Fig1}, we chose $n=1$ in Eq. \eqref{eq:scaling period}. With the initial trapping frequency $\omega_0=2\pi\times897.6\ \textrm{Hz}$ and $\lambda = 0.109$, the round-trip time is $T_1 \approx 2 \ \textrm{ms}$ and the scaling parameter is $\Lambda_{1}^{2} = \exp(2\pi/ \alpha_0) \approx 2.86$. We scan down the trapping frequency to $\omega(t)=\omega_0/\Lambda_{1}^{2}$ at the midpoint $t=T_1/2\approx 1 \ \textrm{ms}$, and then sweep it back according to Eq. \eqref{eq:scaling scan}. We first consider an initial equilibrium state as a reference. As shown in Fig. \ref{Fig1}(a), an initial thermalized atomic cloud reaches the first plateau at the midpoint, and then comes back to the initial size. By contrast, an initial nonequilibrium state is characterized by the breathing-mode oscillation during the trap-holding time [see Figs. \ref{Fig1}(b) and (c)], $\langle r^{2}\rangle(t_{h}) \equiv 
\langle{r^2}\rangle_{0}[1 + A_m\cos(2\omega_{0} t_{h} + \phi_{0})]$, where 
$\langle{r^2}\rangle_{0}$ is the balanced size. The phase is defined by its position on the oscillation, i.e., $\phi \equiv 2\omega_{0} t_{h} + \phi_{0}$. We chose two initial nonequilibrium systems with $\phi_{A}=\pi/2$ and $\phi_{A}=3\pi/2$ as the start points of the trap manipulation, respectively. At the end point, not only does the cloud size return to its initial value, $\langle r^{2}\rangle(t_{B}) \approx \langle r^{2}\rangle(t_{A})$, but so does the dynamical phase, $\phi_{B} \approx \phi_{A}=\pi/2\ (\textrm{or}\ 3\pi/2)$. In Eq. \eqref{scaling-phase} with $n=1$, the phase shift is $\Delta \phi=4\pi$, consistent with the experimental measurement. We note that the phase exists only within the range $[0, 2\pi)$ in the experiment.

Figure \ref{Fig2} shows another experiment for the space-time scale-invariant protocol. Here we chose $n=2$ in Eq. \eqref{eq:scaling period}. For $\lambda = 0.029$, the round-trip time is $T_2 \approx 4 \ \textrm{ms}$. An initially thermalized atomic cloud reaches the second plateau at the midpoint. For the nonequilibrium systems, the sizes return to the initial values at the end point, and the initial phase value is also recovered. 

The above two experiments provide evidence of the reversibility under the space-time scale-invariant protocol. We further observe double reversibility for $n=1$ and $n=2$ (see Figs. S2 and S3 in 
Supplemental Material \cite{Supplemental_Reversibility}), respectively. Under this protocol, the boundary conditions \eqref{eq:C1} and \eqref{eq:C2} require the discreteness of the round-trip time given by Eq. \eqref{eq:scaling period}, which automatically satisfies the phase condition \eqref{eq:C3} of the nonequilibrium system.

\emph{Polynomial STA protocol---} The 
fifth-order polynomial protocol has been previously applied to obtain the STA of a thermalized system  \cite{Beau2016,Deng2018SciAdv,GuerryOdelin2019,Chen2010,Deffner2014,Deng2018,Diao2018}. Here we use this scheme for the reversibility as follows:
\begin{equation}\label{eq:polynomial scan}
b(t)=\\
\begin{cases}
\begin{aligned}[b]
&1+10\eta\bar{t}^3-15\eta\bar{t}^4+6\eta \bar{t}^5,\ 0\leq t \leq T/2, \\[1.2ex]
\end{aligned}\\ 
\begin{aligned}[b]
&1+10\eta\bar{t}_b^3-15\eta\bar{t}_b^4+6\eta\bar{t}_b^5,\ T/2\leq t \leq T,
\end{aligned}
\end{cases}
\end{equation}
where $\bar{t}=2t/T$, $\bar{t}_b=2-2t/T$ and $\eta=b_{ad}-1$. Here, $T$ is the round-trip time, and $b_{ad}=\sqrt{\omega_0/\omega(T/2)}$ is the adiabatic scaling factor at the midpoint $t=T/2$. The trajectory with arbitrary $T$ from the initial state to the end point forms STA paths, i.e., $b(T)=1$ and $\dot{b}(T)=0$.

To satisfy the phase condition \eqref{eq:C3}, the total accumulated phase of the internal oscillation over the round trip should be (see also Supplemental Material \cite{Supplemental_Reversibility}) 
\begin{equation}\label{polynomial phase}
\Delta \phi=2\omega_0 T \int_0^{1}\dfrac{d \bar{t}}{b^2( \bar{t})} = 2n\pi, \ n\in \mathbb{Z}.
\end{equation}
Then the phase condition requires the discreteness of the round-trip time as
\begin{equation} \label{polynomial time}
T_n=\frac{n\pi}{
\displaystyle
\omega_0\int_0^{1}\dfrac{d \bar{t}}{b^2( \bar{t})}},\  n\in \mathbb{Z}.
\end{equation}

Figure \ref{Fig3} shows the experimental measurement for the polynomial STA protocol. Based on the scaling factor $b(t)$ in Eq. \eqref{eq:polynomial scan}, we obtain the trajectory of the trapping frequency $\omega(t)$ from the Ermakov equation \eqref{eq:ermakov}. Here, the initial trapping frequency is $\omega_0 = 2\pi \times 1005(2)\ \mathrm{Hz}$ and is scanned down to $\omega(T/2) = 2\pi \times 500(2)\ \mathrm{Hz}$ at the midpoint, giving the adiabatic scaling factor $b_{\mathrm{ad}} = \sqrt{\omega_0 / \omega(T/2)} \approx \sqrt{2}$. At the start and end points ($A$ and $B$) of the trap manipulation, we obtain the phase of the nonequilibrium state by measuring the oscillation of the breathing mode. When $n$ is set to an integer, such as $n = 6$, the round-trip time is $T_6 =4.16\ \textrm{ms}$ according to Eq. \eqref{polynomial time}. The initially thermalized atomic cloud is reversible and serves as a reference [see Fig. \ref{Fig3}(a)]. For the initial nonequilibrium system with the phase set as $\phi_{A}=0$ [see Fig. \ref{Fig3}(b)], at the end point, not only does the size return to the initial value, $\langle r^{2}\rangle(t_{B}) \approx \langle r^{2}\rangle(t_{A})$, but the phase also recovers, $\phi_{B} \approx \phi_{A}=0$. When $n$ is not an integer, such as $n = 5.5$ and $T_{5.5} =3.80\ \textrm{ms}$, the initially thermalized atomic cloud is still reversible [see Fig. \ref{Fig3}(c)], while the initial nonequilibrium system exhibits obviously irreversible [see Fig. \ref{Fig3}(d)]: the size becomes smaller at the end point, $\langle r^{2}\rangle(t_{B}) < \langle r^{2}\rangle(t_{A})$, and the phase shows much difference, $\phi_{B} \approx \pi$.

In Fig. \ref{Fig3}(e), we schematically show the phase difference, $\Delta\phi=\phi_{B}-\phi_{A}$, as a function of the round-trip time $T$. Due to the phase being constrained in the range $[0, 2\pi)$, $\Delta\phi$ exhibits 
periodic behavior. Figure \ref{Fig3}(f) shows our measurements of $\Delta\phi$ for different $T$, which is consistent with the calculation of Eq. \eqref{polynomial phase}. For arbitrary round-trip time $T$, the thermalized system is reversible. Only when $T$ satisfies Eq. \eqref{polynomial time} with $n$ being an integer, the phase difference $\Delta\phi$ is zero, indicating the reversibility of a nonequilibrium system.

\emph{Conclusion---} We observe the reversibility of a nonequilibrium state in an isotropic unitary Fermi gas. The phase condition that the accumulated dynamic phase of the internal oscillation should be an integer 
multiple of $2\pi$ is required for the reversibility of a 
nonequilibrium state. We find two kinds of pathways, the 
space-time scale-invariant and polynomial STA protocols, to observe the reversibility. The stringent phase condition restricts the round-trip time to be discretized. This work establishes a clear criterion for quantum control of nonequilibrium many-body systems and provides practical pathways for achieving reversibility.

{\it Acknowledgements---} We thank Jing Yang for insightful discussions. This work is supported by the National Key R\&D Program under Grant No. 2022YFA1404102, NSFC (Grants No. 12634010, No. U23A2073, No. 12504312, and No. 12121004), Wuhan city under Grant No. 2024040701010063, and the Luxembourg National Research Fund under Grant Nos. C22/MS/17132060/BeyondKZM and C24/MS/18940482/STAOpen. 

X. W, X. Y and X. X contributed equally to this work.

%

\newpage

\onecolumngrid
\begin{center}
    {\bf \Large Supplemental Material}
\end{center}

\setcounter{secnumdepth}{3} 

\setcounter{equation}{0}

\setcounter{figure}{0}
	
\renewcommand{\thefigure}{S\arabic{figure}}
\renewcommand{\thetable}{S\arabic{table}}
\renewcommand{\theequation}{S\arabic{equation}}

\section{Derivation of the time evolution of the mean-square radius}
\label{sup:cloud_expansion}
For a unitary gas with SO(2,1) symmetry, it is known that the following virial theorem holds \cite{Pitaevskii1997}: 
\begin{equation}
\frac{d^2 \langle{r^2}\rangle(t)}{dt^2} 
+ 
4 \omega^2 \left( t \right)
\langle{r^2}\rangle(t)
= 
\frac{4 E \left( t \right)}{m} 
,
\label{sup:eq:virial}
\end{equation}
where $\langle{r^2}\rangle(t)=\sum_{i = 1}^N \langle{r_i^2}\rangle(t)$ is the mean-square radius, $\omega(t)$ is the trap frequency, $E(t)$ is the total energy of the system, and $m$ is the mass of the ferminonic atom.

From the exact time-evolved state in Eq. (2) in the main text, 
\begin{eqnarray}
\langle{r^2}\rangle(t) 
& = & 
\frac{
\int d \bm{r}_1 
\cdots 
\int d \bm{r}_N \; 
\left( 
\sum_{j = 1}^N 
r^2_j 
\right) 
\left\vert 
\Psi(\{\bm{r}_i\};t) 
\right\vert^2 
}{
\int d \bm{r}_1 
\cdots 
\int d \bm{r}_N \; 
\left\vert 
\Psi(\{\bm{r}_i\};t) 
\right\vert^2 
} 
\nonumber\\
& = & 
\frac{
\int d \bm{r}_1 
\cdots 
\int d \bm{r}_N \; 
\left( 
\sum_{j = 1}^N 
r^2_j 
\right) 
\left\vert 
\Phi\!\left(\!\left\{
\bm{r}_i / b(t)
\right\};\,s(t)\right)
\right\vert^2 
}{
\int d \bm{r}_1 
\cdots 
\int d \bm{r}_N \; 
\left\vert 
\Phi\!\left(\!\left\{
\bm{r}_i / b(t)
\right\};\,s(t)\right)
\right\vert^2 
} 
\nonumber\\
& = & 
b^2 \left( t \right) \langle{r^2}\rangle_\Phi(s(t)) 
, 
\end{eqnarray}
where $\langle{r^2}\rangle_\Phi(s(t))$ is the mean-square radius with respect to the internal state wavefunction $\Phi(\{\bm{r}_i\};s(t))$. 
As $i\hbar\,\partial_s\,\Phi(\{\bm{u}_i\};s) = \hat{H}_0 \,\Phi(\{\bm{u}_i\};s)$ with $\hat{H}_0$ being the initial Hamiltonian with frequency $\omega_{0}$, from Eq. \eqref{sup:eq:virial}, 
\begin{equation}
\frac{d^2 \langle{r^2}\rangle_\Phi(s)}{ds^2} 
+ 
4 \omega^2_0
\langle{r^2}\rangle_\Phi(s) 
= 
\frac{4 E_0}{m} 
, 
\end{equation}
where $E_0 = E(0)$ is constant. 
This leads to 
$\langle{r^2}\rangle_\Phi(s) = \langle{r^2}\rangle_{0} [1 + A_m\cos(2\omega_0 s + \phi_{0})]$ where $\langle{r^2}\rangle_{0} = E_0 / (m \omega^2_0)$, $A_m$, and $\phi_{0}$ are constants.

\section{Trap manipulation for the reversibility } \label{section:I}

\begin{figure}[hpbt]
\includegraphics[width=0.8\textwidth]{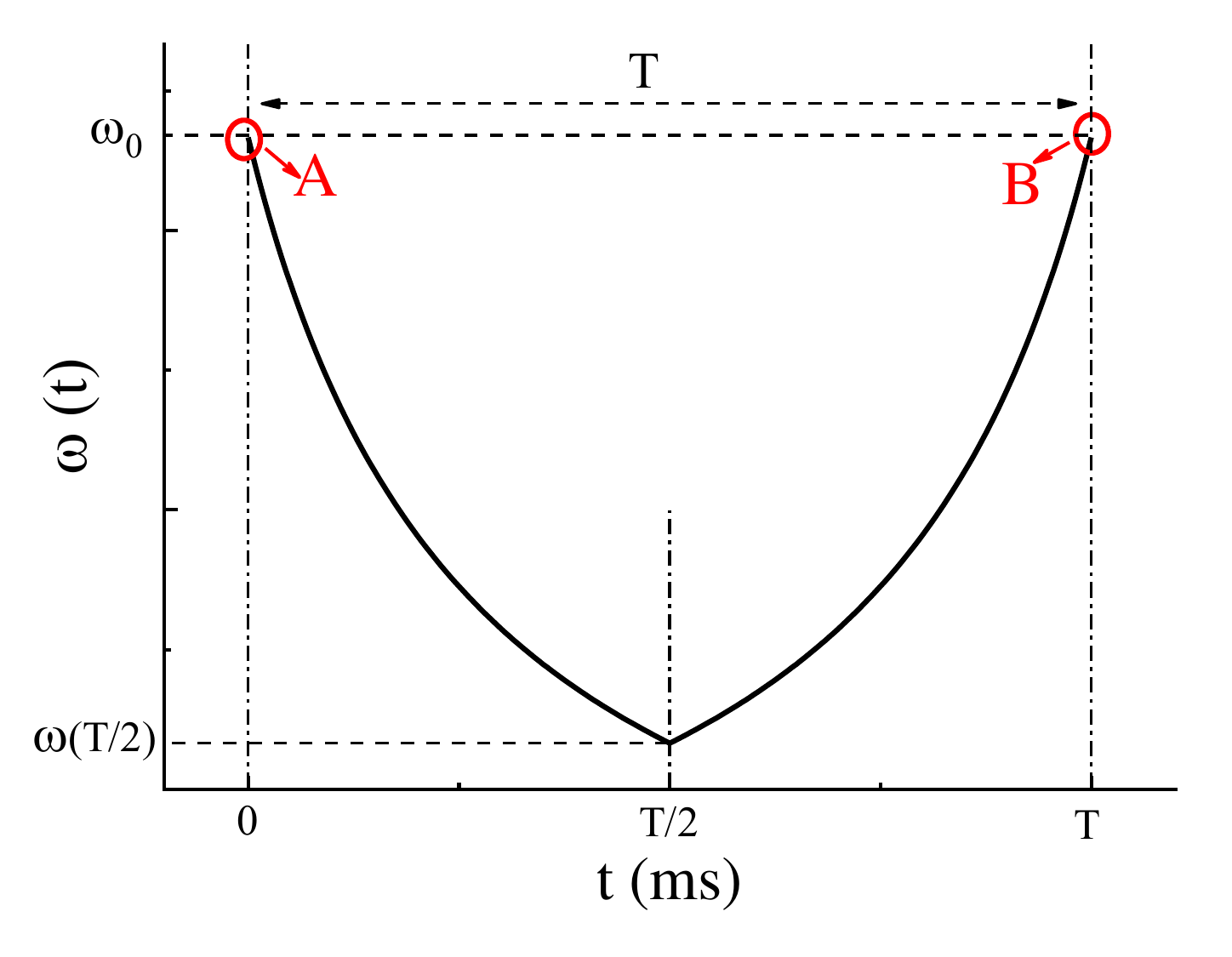}
\caption{Trap manipulation for one reversibility. The initial trapping frequency is $\omega_0$ and the round-trip time of the trap manipulation is $T$. The trapping frequency $\omega(t)$ is scanned down to $\omega(T/2)$ at the midpoint $t=T/2$, and then swept back to the initial value at the end point $t=T$. $A$ denotes the start point of the trap manipulation and $B$ 
the end point. }
\label{FigS1}
\end{figure}

Figure \ref{FigS1} schematically shows how to manipulate the trap to ensure reversibility. The initial trapping frequency is $\omega_0$, and the round-trip time of trap manipulation is $T$. The trapping frequency $\omega(t)$ is scanned down to some value at the midpoint $t=T/2$, and then it is symmetrically swept back to the initial value at the end point of the trap manipulation. Under the space-time scale-invariant protocol, the trajectory of the trapping frequency $\omega(t)$ is presented by Eq. \eqref{eq:scaling scan sup}. Under the polynomial STA protocol, based on the scaling factor $b(t)$ given in Eq. \eqref{eq:polynomial scan sup}, we calculate the trajectory of the trapping frequency $\omega(t)$ from the Ermakov equation.

\section{Double reversibility under the space-time scale-invariant protocol } \label{section:II}

\begin{figure}[hpbt]
\includegraphics[width=0.8\textwidth]{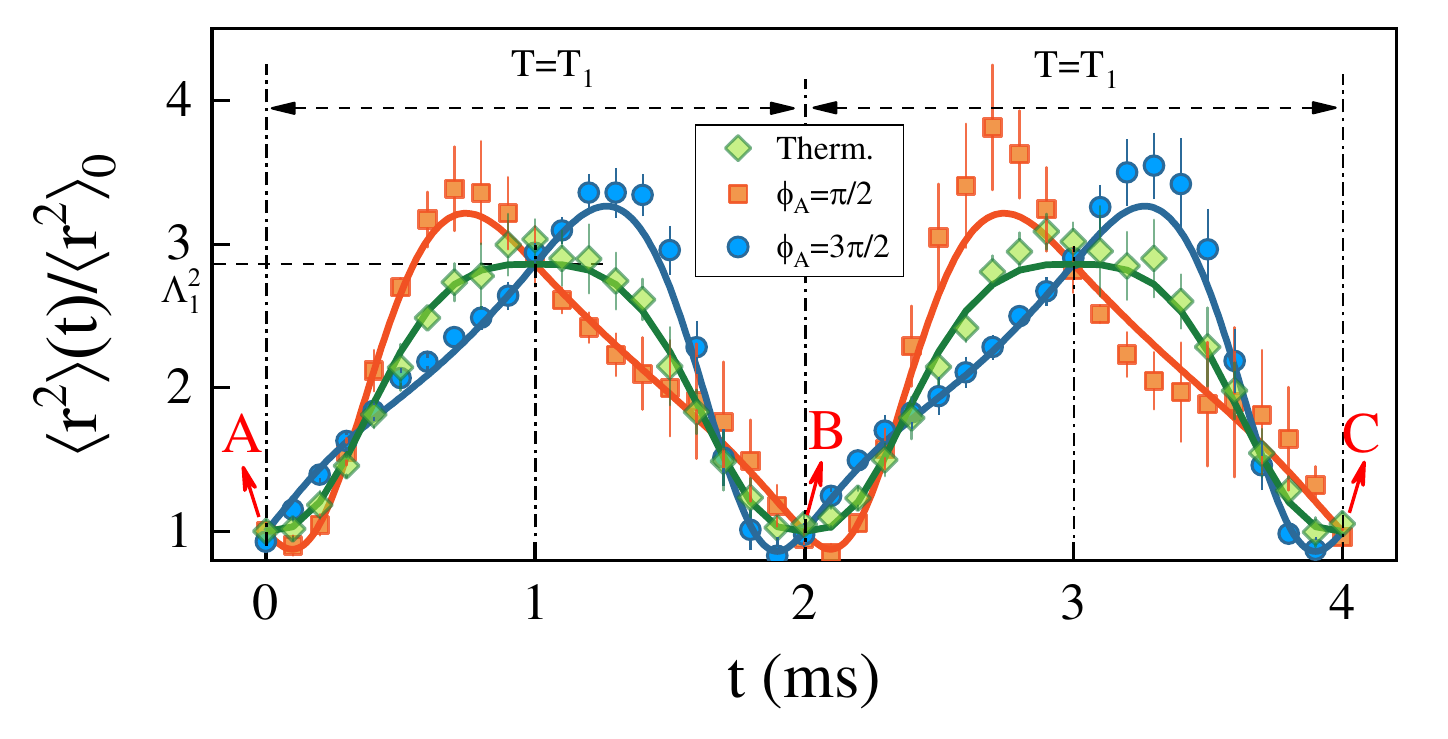}
\caption{Double reversibility with $n = 1$ under the space-time scale-invariant protocol. Label $A$ denotes the start point of the trap scan, $B$ refers to the first reversibility point after the first round trip of the trap ramp, and $C$ marks the second reversibility point. With $\lambda = 0.109$, the round-trip time is $T=T_1 \approx 2 \ \textrm{ms}$. Normalized atomic cloud sizes $\langle{r^2}\rangle(t)/\langle{r^2}\rangle_{0}$ are shown during the trap scan. An initially thermalized atomic cloud (green diamonds) is shown as a reference, and at the midpoints $t=1\ \textrm{or}\ 3\ \textrm{ms}$, it reaches the first plateau with $\Lambda_{1}^{2} = 2.86$. The phases of the two initial nonequilibrium systems are set as $\phi_{A}=\pi/2$ (orange squares) and $\phi_{A}=3\pi/2$ (blue circles), respectively. At $B$ and $C$ points, the atomic cloud sizes return to their initial values for the thermalized and nonequilibrium systems, i.e., $\langle{r^2}\rangle(t_B) \approx \langle{r^2}\rangle(t_C) \approx \langle{r^2}\rangle(t_A)$. The solid curves are the calculations of Eq. 
\SHS{(4) }
in the main text. Error bars represent the standard deviation of several measurements.}
\label{FigS2}
\end{figure}

In Fig. \ref{FigS2}, we observe double reversibility with $n = 1$ under the space-time scale-invariant protocol. The experimental conditions are the same as those in Fig. 1 of the main text. With $\lambda = 0.109$, the round-trip time is $T=T_1 \approx 2 \ \textrm{ms}$. Here we scan the trapping frequency with two periods. An initially thermalized atomic cloud reaches the first plateau at the midpoint, and then come back to the initial size, serving as a reference. The phases of the two initial nonequilibrium systems are set as $\phi_{A}=\pi/2$ and $\phi_{A}=3\pi/2$, respectively. At the two end points of the trap scan, the atomic cloud sizes return to their initial values for the thermalized and nonequilibrium systems, i.e.,  $\langle{r^2}\rangle(t_B) \approx \langle{r^2}\rangle(t_C) \approx \langle{r^2}\rangle(t_A)$. By measuring the breathing-mode oscillation, the phases also recover, $\phi_{B} \approx \phi_{C} \approx \phi_{A}=\pi/2\ (\textrm{or}\ 3\pi/2)$. 

\begin{figure}[hpbt]
\includegraphics[width=0.8\textwidth]{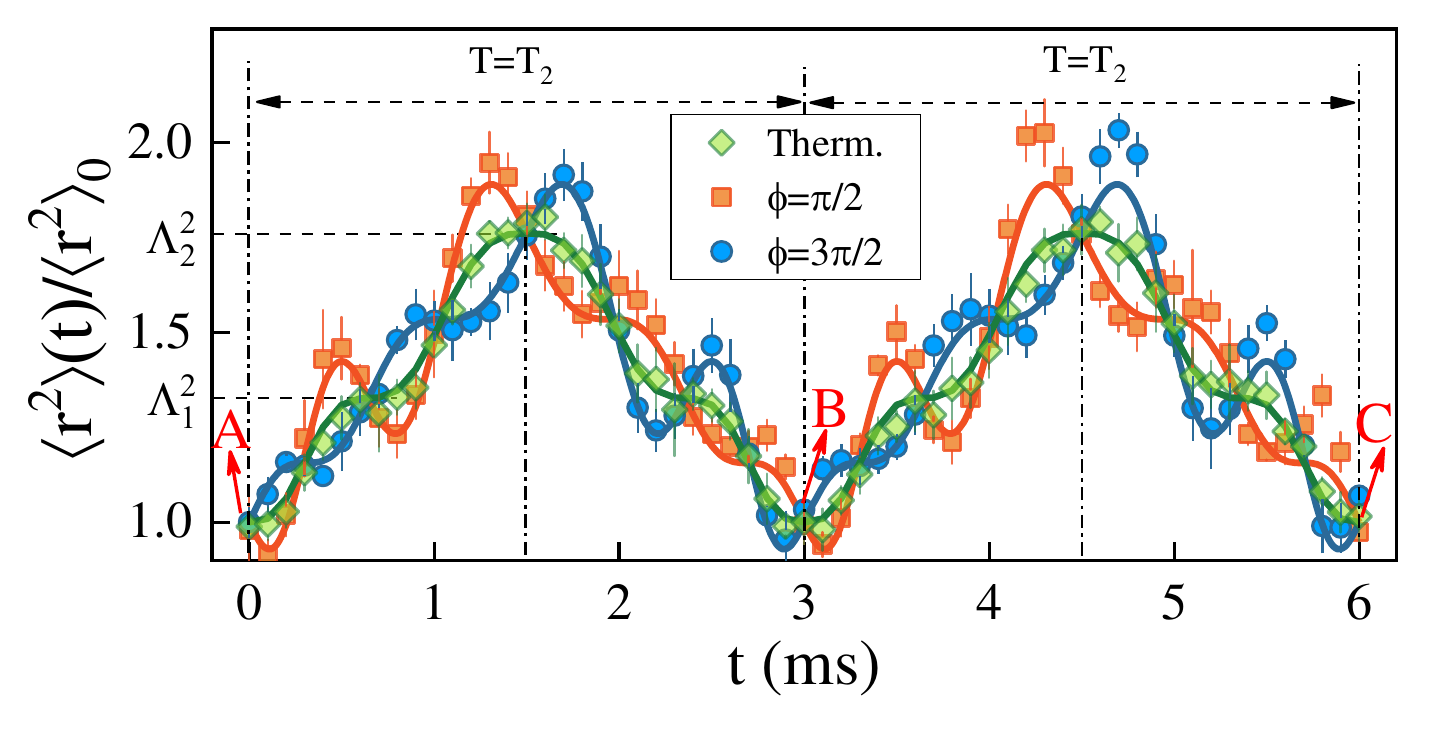}
\caption{Double reversibility with $n = 2$ under the space-time scale-invariant protocol. $A$ denotes the start point of the trap scan, $B$ 
the first reversibility point after the first round trip of the trap ramp, and $C$ 
the second reversibility point. Normalized atomic cloud sizes $\langle{r^2}\rangle(t)/\langle{r^2}\rangle_{0}$ are shown during the trap scan. With $\lambda = 0.008$, the round-trip time is $T=T_2 \approx 3 \ \textrm{ms}$. An initially thermalized atomic cloud (green diamonds) is shown as a reference, and at the midpoints $t=1.5\ \textrm{or}\ 4.5\ \textrm{ms}$, it reaches the second plateau with $\Lambda_{2}^{2} = 1.76$. The phases of the two initial nonequilibrium systems are set as $\phi_{A}=\pi/2$ (orange squares) and $\phi_{A}=3\pi/2$ (blue circles), respectively. At $B$ and $C$ points, the atomic cloud sizes return to their initial values for the thermalized and nonequilibrium systems, i.e.,  $\langle{r^2}\rangle(t_B) \approx \langle{r^2}\rangle(t_C) \approx \langle{r^2}\rangle(t_A)$. The solid curves are the calculations of Eq. 
(4) in the main text. Error bars represent the standard deviation of several measurements. }
\label{FigS3}
\end{figure}

In Fig. \ref{FigS3}, we observe double reversibility with $n = 2$ under the space-time scale-invariant protocol.
At variance with Fig. 2 of the main text, the scan parameter is $\lambda = 0.008$ and the round-trip time is $T=T_2 \approx 3 \ \textrm{ms}$. We scan the trapping frequency for two periods. An initially thermalized atomic cloud reaches the second plateau at the midpoint, and then 
comes back to the initial size, serving as a reference. The phases of the two initial nonequilibrium systems are set as $\phi_{A}=\pi/2$ and $\phi_{A}=3\pi/2$, respectively. At the two end points of the trap scan, the atomic cloud sizes return to their initial values for the thermalized and nonequilibrium systems, i.e.,  $\langle{r^2}\rangle(t_B) \approx \langle{r^2}\rangle(t_C) \approx \langle{r^2}\rangle(t_A)$. By measuring the breathing-mode oscillation, the phases also recover, $\phi_{B} \approx \phi_{C} \approx \phi_{A}=\pi/2\ (\textrm{or}\ 3\pi/2)$. 

\section{Calculation of the phase condition under the space-time scale-invariant protocol} \label{section:III}
Consider the variation of the trapping frequency 
discussed in 
the main text,  
\begin{equation} \label{eq:scaling scan sup}
\omega(t) = \begin{cases}
\dfrac{1}{\sqrt{\lambda}(t_0+t)}, &  0\leq t \leq T/2, \\[1.2ex]
\dfrac{1}{\sqrt{\lambda}(T+t_0-t)}, & T/2 \leq t \leq T, 
\end{cases}
\end{equation}
where $\omega_0=1/(\sqrt{\lambda}t_0)$ is the initial trapping frequency, $T$ is the round-trip time, and $\lambda$ characterizes the speed of the trap ramp. 

To make the system reach 
the $n$-th plateau at $t=T_n/2$ with $T = T_n$, 
the following condition should be satisfied: 
\begin{equation} \label{eq:scaling period sup}
T_{n}/2=t_0(e^{2n\pi/\alpha_0}-1), \ n\in \mathbb{Z},
\end{equation}
where $\alpha_0=2\sqrt{1/\lambda-1/4}$.

For the system satisfying Eqs. \eqref{eq:scaling scan sup} and \eqref{eq:scaling period sup}, the Ermakov equation has the solution $b(t)$ with $b(0)=b(T) = 1$ and $\dot{b}(0)=\dot{b}(T) = 0$, which is given by
\begin{equation} \label{eq:scaling factor}
b^2(t) = 
\begin{cases}
\begin{aligned}[t]
&\dfrac{t_0+t}{t_0\sin^2\varphi}\left[1-\cos\varphi\cdot\cos(
\alpha_0
\ln\frac{t_0+t}{t_0}+\varphi)\right],  \qquad  0\leq t \leq T/2, 
\end{aligned} \\[1.5ex]
\begin{aligned}[t]
&\dfrac{T+t_0-t}{t_0\sin^2\varphi}\left[1-\cos\varphi\cdot\cos(
\alpha_0
\ln\frac{T+t_0-t}{t_0}+\varphi)\right], \qquad  T/2 \leq t \leq T,
\end{aligned}
\end{cases}
\end{equation}
where $\varphi=-\arctan \alpha_0$.
According to scaling factor in Eq. \eqref{eq:scaling factor}, the rescaled time from $t=0$ to $t=T$ is
\begin{equation}\label{eq:scaling rescaled time}
s(T)=\int_{0}^{T/2}\frac{dt}{b^2(t)}+\int_{T/2}^{T}\frac{dt}{b^2(t)}=\frac{4n\pi t_0}{\sqrt{1+
\alpha_0^2
}}=\frac{2n\pi}{\omega_0},\ n\in \mathbb{Z}.
\end{equation}
Then the accumulated dynamic phase of the internal oscillation during the round-trip time $T$ is
\begin{equation} \label{scaling-phase-sup}
\Delta \phi =2\omega_0 s(T)=4n\pi,\   n\in \mathbb{Z},
\end{equation}
which automatically satisfies the phase condition.

\section{Calculation of the phase condition under the polynomial STA protocol} \label{section:IV}

The fifth-order polynomial protocol has been previously applied to obtain the STA of a thermerlized system \cite{Beau2016,Deng2018SciAdv,GuerryOdelin2019,Chen2010,Deffner2014,Deng2018,Diao2018}. Here we use this scheme for the reversibility as follows:
\begin{equation}\label{eq:polynomial scan sup}
b(t)=\\
\begin{cases}
\begin{aligned}[b]
&1+10\eta\bar{t}^3-15\eta\bar{t}^4+6\eta \bar{t}^5,\ \ 0\leq t \leq T/2, \\[1.2ex]
\end{aligned}\\ 
\begin{aligned}[b]
&1+10\eta\bar{t}_b^3-15\eta\bar{t}_b^4+6\eta\bar{t}_b^5,\ \ T/2\leq t \leq T,
\end{aligned}
\end{cases}
\end{equation}
where $\bar{t}=2t/T$, $\bar{t}_b=2-2t/T$, and $\eta=b_{ad}-1$. Here, $T$ is the round-trip time of the trap ramp, and $b_{ad}=\sqrt{\omega_0/\omega(T/2)}$ is the adiabatic scaling factor at the midpoint $t=T/2$. The trajectory with arbitrary $T$ from the initial state to the end point forms STA paths, i.e., $b(T)=1$ and $\dot{b}(T)=0$.

Along the trajectory in Eq. \eqref{eq:polynomial scan sup}, the rescaled time is
\begin{equation} \label{eq:polynomial rescaled time}
\begin{split}
s(T) &= \int_0^{T/2}\frac{1}{b^2(t)}dt+\int_{T/2}^{T}\frac{1}{b^2(t)}dt  = T\int_0^{1}\dfrac{1}{b^2(\bar{t})}d \bar{t}.
\end{split}
\end{equation}
To satisfy the phase condition, the total accumulated phase of the internal oscillation over the round trip should be
\begin{equation}\label{polynomial phase sup}
\Delta \phi=2\omega_0 T \int_0^{1}\dfrac{1}{b^2( \bar{t})}d \bar{t} = 2n\pi, \ n\in \mathbb{Z}.
\end{equation}
Then the phase condition requires the discreteness of the 
round-trip 
time as
\begin{equation} \label{polynomial time sup}
T_n=\frac{n\pi}{
\displaystyle
\omega_0\int_0^{1}\dfrac{1}{b^2( \bar{t})}d \bar{t}},\  n\in \mathbb{Z}.
\end{equation}

\end{document}